\documentclass[showpacs,superscriptaddress,floatfix,twocolumn]{revtex4-1}
\usepackage{amsmath,amssymb,amsfonts}
\usepackage{graphicx,epsfig}
\usepackage{subfigure}
\usepackage{xcolor}
\usepackage{times}
\newcommand{\braket}[1]{\left\langle{#1}\right\rangle}

\newcommand{\iD}{\overline{\Delta_3}}
\newcommand{\iPk}{\overline{P_k^\delta}}

\begin{document}
\title{Accuracy and precision of the estimation of the number of missing levels in chaotic spectra using long-range correlations}
\author{I. Casal}
\email{ignacio.casal@uam.es}
\affiliation{Grupo de F\'{\i}sica Nuclear, Departamento de Estructura de la Materia, F\'{\i}sica T\'ermica y Electr\'onica, Facultad de Ciencias F\'{\i}sicas, Universidad Complutense de Madrid and IPARCOS, CEI Moncloa, Madrid, 28040, Spain}
\author{L. Mu\~noz}
\email{lmunoz@ucm.es}
\affiliation{Grupo de F\'{\i}sica Nuclear, Departamento de Estructura de la Materia, F\'{\i}sica T\'ermica y Electr\'onica, Facultad de Ciencias F\'{\i}sicas, Universidad Complutense de Madrid and IPARCOS, CEI Moncloa, Madrid, 28040, Spain}
\author{R. A. Molina}
\email{rafael.molina@csic.es}
\affiliation{Instituto de Estructura de la Materia, IEM-CSIC, Serrano 123, Madrid, E-28006, Spain}

\begin{abstract}
We study the accuracy and precision for estimating the fraction of observed levels $\varphi$ in quantum chaotic spectra through long-range correlations. We focus on the main statistics where theoretical formulas for the fraction of missing levels have been derived, the $\Delta_3$ of Dyson and Mehta and the power spectrum of the $\delta_n$ statistic. We use Monte Carlo simulations of the spectra from the diagonalization of Gaussian Orthogonal Ensemble matrices with a definite number of levels randomly taken out to fit the formulas and calculate the distribution of the estimators for different sizes of the spectrum and values of $\varphi$. 
A proper averaging of the power spectrum of the $\delta_n$ statistic needs to be performed for avoiding systematic errors in the estimation. Once the proper averaging is made the estimation of the fraction of observed levels has quite good accuracy for the two methods even for the lowest dimensions we consider $d=100$. However, the precision is generally better for the estimation using the power spectrum of the $\delta_n$ as compared to the estimation using the $\Delta_3$ statistic. This difference is clearly bigger for larger dimensions. 
Our results show that a careful analysis of the value of the fit in view of the ensemble distribution of the estimations is mandatory for understanding its actual significance and give a realistic error interval.
\end{abstract}


\maketitle

\section{Introduction}

Statistical analysis of spectra has become a very useful tool in Physics. It already started in the 50-60's of the past century in the field of Nuclear Physics when Wigner, Dyson, Gaudin, Mehta and others developed and applied the Random Matrix Theory (RMT) to nuclear spectra \cite{Wigner1955,Wigner1957,Wigner1958,Wigner1967,Mehta1960,Gaudin1961,Dyson1962a,Dyson1962b,Dyson1963,Mehta1963,Flores1981,Guhr1998}. 
One of the main results arrived
in 1982 when Haq, Pandey and Bohigas analyzed the spectral fluctuations of an ensemble of 1407 experimentally identified neutron and proton $J^{\pi}=1/2^+$ resonances just above the one-nucleon emission threshold (the Nuclear Data Ensemble, NDE) \cite{Haq1982}. They showed an impressive agreement with the prediction of the Gaussian Orthogonal Ensemble (GOE) of RMT with very high statistical significance.
The results are certainly striking considering the matrix models from RMT are constructed with random numbers and the prediction is parameter-free.
Two years later, Bohigas, Giannoni and Schmit applied RMT to study the spectral fluctuations of a Sinai quantum billiard, whose classical analogue shows chaotic dynamics, obtaining also a very good agreement with the GOE \cite{Bohigas1984}. In view of this result, 
they established as a conjecture that the spectral fluctuations of time-reversal-invariant systems whose classical analogs are chaotic are the same as those predicted by GOE. Known from then as the BGS conjecture, it has been supported by many other results in different physical systems along the years, and it was finally proved semiclassically by Heusler {\it et al.} in 2007 \cite{Mueller2004}. On the other hand, for quantum systems with an integrable classical analogue Berry and Tabor showed in 1977 that their spectral fluctuations are identical to those of a sequence of uncorrelated random numbers and follow Poisson statistics \cite{Berry1977}.

Once the classification
depending on the underlying dynamics of the classical system is made, the way is paved to extend it to all quantum systems with or without a classical analogue. Thus, a quantum system is said to be chaotic when its spectral fluctuations are well described by RMT and it is said to be regular when they follow Poisson statistics. Spectral fluctuations of quantum systems are universal as they do not depend on the particular properties of each system but only on the more general properties of the dynamics, whether it is regular or chaotic. However, this universality can only be argued for the two extremes, systems intermediate between chaos and regularity lack this universal characterization.

To measure spectral fluctuations in RMT a certain statistic is defined which can be calculated from the sequence of levels of a quantum spectrum. The statistics are usually divided on whether they measure short range correlations between neighboring levels like the nearest-neighbor spacing distribution $P(s)$ and the distribution of ratios between neighboring spacings \cite{Flores1981,Oganesyan2007} or long-range correlations between distant levels like the $\Delta_3$ of Dyson and Mehta \cite{Mehta1963} or the power spectrum of the $\delta_n$ \cite{Relano2002}. In order to compare experimental spectra with RMT results, complete sequences of levels without mixed symmetries are needed. In actual experiments missing levels are sometimes unavoidable, for example, if the wave function has a node in the location of the antenna in a microwave billiard experiment the level cannot be measured. There are then a number of works that have worked out how the correlations in the spectral fluctuations depend on the fraction of observed levels for chaotic quantum systems \cite{Bohigas2004,Molina2007,Mulhall2011}. In general, as the fraction of missing levels increases the correlations in the spectra diminish and the spectral statistics become closer to the Poisson case. Assuming the chaoticity of the quantum system, the fraction of missing levels in the spectral sequence can be estimated from the spectral statistics. This way of estimating the number of missing levels in experimental spectra has been succesfully applied for different systems \cite{MurPetit2015,Lawniczak2018}. However, very few attempts have been made to address the quality of the different estimators and to make a comparison between them. 
This is specially relevant as systems intermediate between regularity and chaos as well as systems with a superposition of sequences with different symmetries also present correlations between the RMT result and Poisson.
It is therefore necessary to establish how to correctly use each statistic for each kind of transition in order to avoid misleading conclusions on the origin of the intermediate behavior of a certain system. 

The purpose of this work is to contribute to the systematization of the correct use of the available statistics to estimate the fraction of missing levels in spectra. In particular, we focus on the $\Delta_3$ and the $P_k^{\delta}$ statistics, which measure long-range correlations between levels. For both, there exist theoretical formulas
for the transition as a function of one parameter, the fraction of observed levels, ($\varphi \in [0,1]$).
However, the theoretical ensemble standard deviation has not been calculated before. The square root of the ensemble standard deviation of $\varphi$ is a measure of the error in the estimation of $\varphi$ from a fit to the theoretical formulas and should be as small as possible \cite{Mehta_Book}. It will depend on the particular statistic used but also on the value of $\varphi$ and the size of the analyzed level sequence. This error is very different from the simple numerical error of a non-linear squares fit and is the relevant one in RMT studies, as we show in this analysis.
In this work we resort to a numerical calculation of the full distribution of $\varphi$. This will allow us to study the accuracy, related to the correctness of the theoretical formulas, and the precision, related to ensemble standard deviation, of the estimation of the fraction observed levels, depending on its value, the size of the spectrum and the method of calculation.

The paper is organized as follows: In Sec. \ref{sec:Stat} we describe the tools we use for the Montecarlo simulations of RMT spectra with missing levels and the statistics we use for the analysis of long-range correlations and the estimation of the fraction of observed levels. In Sec. \ref{sec:results} we show the main results of the paper, {\em i.e.} the distribution of 
the estimations of the fraction of observed levels depending on the different methods used. The tables and figures of this section summarize the results for the precision and accuracy of the fit of the fraction of observed levels depending on the size of the level sequence. In Sec. \ref{sec:conclusions} we summarize the main conclusions of the work.

\section{Statistical analysis of RMT spectra with missing levels}
\label{sec:Stat}

For analyzing the accuracy and precision of the estimation of the fraction of missing levels we need to construct ensembles of chaotic spectra with different fractions of observed levels $\varphi$. We describe below the procedure we have used focusing on GOE spectra.

The GOE is an ensemble of matrices defined in RMT to represent hamiltonian matrices with time-reversal symmetry and spin rotation symmetry, that is, the vast majority of physical hamiltonians. Thus, it is the most used ensemble to compare with experimental results and perform numerical and theoretical studies. The GOE is composed of real symmetric matrices invariant under orthogonal transformations. The matrix elements are random numbers from Gaussian distributions and together with the GUE (Gaussian Unitary Ensemble, for hamiltonians without time-reversal symmetry) and the GSE (Gaussian Simplectic Ensemble, for hamiltonians with time-reversal symmetry but no spin rotational symmetry) they constitute the three classical ensembles of RMT.

First of all, we generate a sample of matrices belonging to the GOE and diagonalize each of them to obtain the spectrum. We then randomly eliminate the necessary amount of levels to obtain a spectrum with a determined fraction of observed levels $\varphi$. 

After that, the first necessary step prior to the statistical analysis is the so-called {\it unfolding}, that is, to separate the spectral fluctuations, which are the object of study, from the secular behavior of the level density. To perform the unfolding one needs to assume that the density of states $g(E)$ can be separated into a smooth part $\overline{g}(E)$ and a fluctuating part $\widetilde{g}(E)$,
\begin{equation}
  g(E) = \overline{g}(E) + \widetilde{g}(E)
\end{equation}
The standard procedure by which $\overline{g}(E)$ is removed consists in mapping the actual energy levels $\{E_i\}_{i=1,\ldots,d}$ into new dimensionless levels $\{\varepsilon_i\}_{i=1,\ldots,d}$ whose mean level density is constant. Here $d$ stands for the size of the spectrum. This can be done by means of the following transformation:
\begin{equation}
  \varepsilon_i = \overline{N}(E_i), \hspace{1cm} i=1,\ldots,d,
\end{equation}
where $\overline{N}(E)$ is the smooth part of the accumulated level density $N(E) = \int_{-\infty}^E dE' g(E')$, which gives the number of levels up to energy $E$. The transformed level density $\rho(\varepsilon)$ in the new energy variable $\varepsilon$ is such that $\overline{\rho}(\varepsilon) = 1$, as required.

A correct unfolding is crucial in order to properly analyze the spectral fluctuations. Ideally the smooth part of the level density, that is, the mean level density should be perfectly known. For physical systems this is not always the case. 
In many relevant cases only a very general fit, like a fit to a polynomial function of the energy, can be done. A polynomial fit is a very illustrative example to understand the consequences of the possible mistakes which can be made. If the degree of the polynomial is too low the fit cannot accurately describe $\overline{g}(E)$ completely and if it is too high the fit can be too good in the sense that it would also partially include the fluctuations and not only the smooth part. Both type of errors lead to misleading conclusions in the spectral analysis \cite{Gomez2002}.

In this work we deal with spectra from the GOE, whose level density is perfectly known and, moreover, the change in the level density when there are missing levels is straightforward. The mean level density of the GOE is known as Wigner's semicircle law and is given by:
\begin{equation}
\label{ec:semicirculo}
\overline{\rho}(E) =
\begin{cases} 
  \frac{A}{\pi}\sqrt{\frac{2d}{A} - E^2} & |E| < \sqrt{\frac{2d}{A}} \\
  0       & |E| > \sqrt{\frac{2d}{A}}
  \end{cases}
\end{equation}
where $A$ is a constant. 
As the level density represents the number of energy levels per unit energy, we only need to multiply by a factor $\varphi$ to have the level density of a spectrum with a fraction $\varphi$ of observed levels.

Thus, the mean accumulated level density we use here for the unfolding is:
\begin{align}
\overline{N}(E) &= \frac{\varphi}{2}\left(\frac{A}{\pi}\left[E\sqrt{\frac{2d}{A} - E^2} \right. \right. \nonumber\\
 &+ \left. \left.  \frac{2d}{A}\arcsin\left(\frac{E}{\sqrt{2d/A}}\right)\right] + d\right)
\label{ec:NE_func}
\end{align}

Once we have an ensemble of unfolded spectra with a certain fraction $\varphi$ of observed levels, we can proceed to calculate the two statistics we focus on: $\Delta_3$ and $P_k^{\delta}$.

The $\Delta_3$ is one of the most widely used long-range statistics and was defined originally by Dyson and Mehta \cite{Dyson1963,Mehta1963}. It is defined in an interval $[\varepsilon, \varepsilon + L]$ of the unfolded spectrum as
\begin{equation}
\Delta_3(L) = \braket{\min_{A,B} \frac{1}{L} \int_{\varepsilon}^{\varepsilon+L} d\varepsilon' \left[N(\varepsilon')-A\varepsilon'-B\right]^2},
\label{eq:delta3}
\end{equation}
where the angle brackets denote the spectral average over the values of $\varepsilon$, the location of the window of $L$ levels within the spectrum. The value of $\Delta_3$ is independent of the position of $\varepsilon$ and that is why it makes sense to calculate an average over intervals of length $L$ along the spectrum.

The $\Delta_3$ is a measure of the {\it spectral rigidity} or of how ``ordered'' is the spectrum in the following sense: The nearer to an equally spaced spectrum the smaller the value of $\Delta_3$ is. For an exactly equally spaced spectrum it has a constant value $\iD = 1/12$. The opposite is a regular system with Poisson statistics where there are no correlations between levels and it grows linearly in this case, $\iD = L/15$. GOE spectra are more {\it rigid} and the growth with $L$ is logarithmic. The result cannot be computed analytically but it is possible to obtain an asymptotic expression valid for large $L$:

\begin{equation}
\iD(L) = \frac{1}{\pi^2}\left[\log(2\pi L) + \gamma - \frac{5}{4}\right] - \frac{1}{8} + \mathcal{O}(L^{-1})
\label{eq:GOEdelta3}
\end{equation}

This equation is sufficient for most purposes as the difference between this asymptotic formula and the numerical result is negligible for $L\gtrsim 6$. The bar over $\Delta_3$ denotes average over the ensemble, as these are the theoretical predictions which can be supplied by RMT.
For the practical calculation of $\iD$ we have implemented the prescription described in Ref. \cite{Mulhall2011}. It must be noted that in order to have a meaningful spectral average it is important to calculate $\Delta_3(L)$ through independent intervals of length $L$ in the spectrum. The number of such intervals limits the maximum value of $L$ for which the $\Delta_3(L)$ can be calculated \cite{Bohigas1975}. In this work we have chosen to limit the calculation to $L=d/3.5$ in order to average over at least three fully independent intervals. 

The $P_k^{\delta}$ statistic is the power spectrum of the $\delta_n$ statistic which is defined in terms of the unfolded energy levels as
\begin{equation}
\delta_n = \varepsilon_{n+1} - \varepsilon_1 - n, \hspace{1.5cm} n = 1,\ldots,d-1
\label{eq:deltan}
\end{equation}
and it represents the deviation of the excitation energy of the $(n+1)th$ unfolded level from its mean value $n$. Moreover, if we appropriately shift the ground state of the system, we can write
\begin{equation}
  \delta_n = -\widetilde{N}(\varepsilon_{n+1}),
\end{equation}
that is, the accumulated level density fluctuations at $\varepsilon = \varepsilon_{n+1}$.

The $\delta_n$ statistic was regarded in \cite{Relano2002} from a new point of view, considering its formal similarity with a discrete time series, resulting in a new long-range statistic to characterize quantum chaos. One of the most common numerical techniques used in time series analysis is the calculation of the power spectrum, the square modulus of the Fourier transform:

\begin{equation}
P^\delta_k \equiv |\hat{\delta}_k|^2 = \frac{1}{N}\left|\displaystyle\sum_{n = 1}^N\delta_n\exp\left(\frac{-i2\pi nk}{N}\right)\right|^2
\label{ec:Pk}
\end{equation}

The three classical ensembles of RMT (GOE, GUE and GSE) showed $P_k^{\delta} \propto 1/k$, while on the other hand, Poisson spectra showed $P_k^{\delta} \propto 1/k^2$. In view of these results it was conjectured that chaotic quantum systems are characterized by $1/f$ noise, whereas integrable ones exhibit $1/f^2$ noise \cite{Relano2002}. 
The analogy with time series also provides a consistent interpretation of spectral rigidity in terms of antipersistence. In a time series, antipersistence means that an increasing or decreasing trend in the past increases the probability of the opposite trend in the future. Fluctuations in an unfolded spectrum are the deviations from an equally spaced spectrum with spacings between levels equal to 1. Thus, if the signal $\delta_n$, viewed as a time series, is very antipersistent, the corresponding energy spectrum is very rigid. This is the case of RMT and chaotic spectra. 

The $P_k^{\delta}$ statistic has been used to study numerical spectra from theoretical models like RMT ensembles \cite{Faleiro2004,Relano2004,Santhanam2005,Male2007}, the nuclear shell-model \cite{Relano2002}, quantum billiards \cite{Gomez2005}, the quartic oscillator and the kicked top \cite{Santhanam2005} to study the chaotic dynamics of quantum models or the order-chaos transition in mixed systems. Microwave billiards \cite{Lawniczak2018, Bialous2019}, microwave networks \cite{Dietz2017, Bialous2016} or molecules \cite{MurPetit2015} are some of the experimental spectra recently analyzed using $P_k^\delta$ to address issues such as chaoticity, particular non-universal features in the spectra or missing levels. 

For our analysis we use the following expressions of $\Delta_3(L)$ \cite{Bohigas2004,Mulhall2011} and $P_k^{\delta}$ \cite{Molina2007} for a GOE spectrum with only a fraction $\varphi$ of observed levels:
\begin{align}
\label{ec:D3phi}
\iD(L;\varphi) &=  \frac{\varphi^2}{\pi^2}\left[\log\left( \frac{2\pi L}{\varphi}\right) + \gamma - \frac{5}{4}-\frac{\pi^2}{8} \right] \nonumber\\ &+ (1 - \varphi)\frac{L}{15} \\
\label{ec:Pkphi}
\iPk(\varphi) &= \frac{N^2 \varphi}{4\pi^2}\left[\frac{K(\varphi\frac{k}{N}) - 1}{k^2} + \frac{K\left(\varphi\frac{N - k}{N}\right) - 1}{(N - k)^2}\right] \nonumber \\ &+ \frac{1}{4\sin^2\left(\frac{\pi k}{N}\right)} - \frac{\varphi^2}{12}
\end{align}

\begin{figure}[h]
\includegraphics[width=0.4\textwidth]{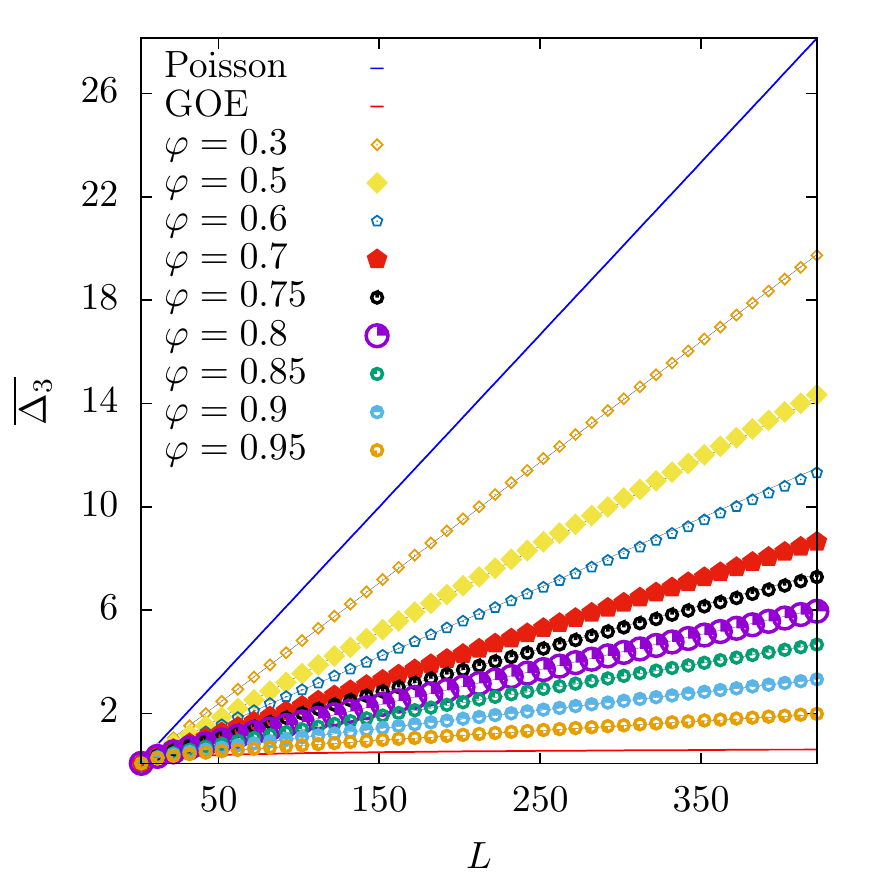}
\caption{$\iD(L)$ vs. $L$ for different values of the number of missing levels. The curves are the ensemble average of ensembles of 1500 matrices of dimension $d=1500$ where a fraction of levels have been randomly taken out leaving a fraction $\varphi$ of observed levels. Results are compared with the theoretical formula \eqref{ec:D3phi}.}
\label{fig:form_ensembles_D3phi}
\end{figure}

\begin{figure}[h]
\includegraphics[width=0.4\textwidth]{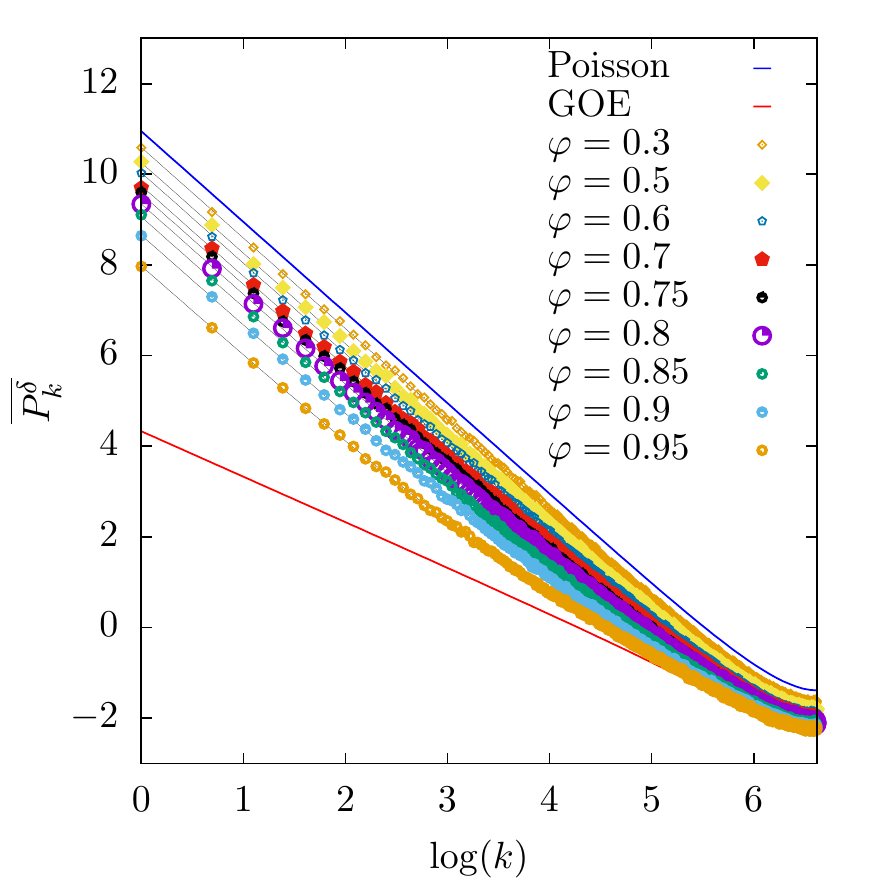}
\caption{Value of $\iPk$ vs. $k$ in logarithmic scale for the same data as Fig. \ref{fig:form_ensembles_D3phi} and comparison with the theoretical formula \eqref{ec:Pkphi}.}
\label{fig:form_ensembles_Pkphi}
\end{figure}

The formula for $\iPk(\varphi)$ uses the so-called form factor approximation that only takes into account correlations between two levels. Analytical results for the GUE have recently obtained expressions beyond the form factor approximation that take into account correlations to all orders \cite{Riser2017,Riser2020,Corps2019}. Unfortunately, similar analytical formulas for the GOE have not been obtained yet. The differences are evident only at large values of $k$ close to the limit $d/2$. For our purposes, the formulas coming from the form factor approximation are good enough for estimating the fraction of observed levels $\varphi$ as we show below.

The quality of these formulas for describing ensemble averages of chaotic spectra with missing levels has been shown before \cite{Bohigas2004,Mulhall2011,Molina2007}. However, in order to be self-contained, we show in Figs. \ref{fig:form_ensembles_D3phi} and \ref{fig:form_ensembles_Pkphi} the curves for several values of $\varphi$ to compare the behavior of the two statistics as $\varphi$ varies from the GOE prediction ($\varphi = 1$) to the Poisson case ($\varphi = 0$) and compare with the results of averages
over the corresponding ensembles of 1500 members each for dimension $d=1500$ 
as a check of the correctness of the theoretical predictions.

Similar formulas have been obtained for another widely used statistic to study long-range correlations, the $\Sigma^2$ \cite{Bohigas2004}.
The $\Delta_3$ is an integrated version of the $\Sigma^2$ and, thus, much smoother. Although, the quality of the formulas for the ensemble average of $\Delta_3$ and $\Sigma^2$ is very similar, the variance within the ensemble is generally much higher for the $\Sigma^2$ than for the $\Delta_3$. The latter is, thus, better suited for estimating the number of missing levels and has been used in past and recent works with this aim \cite{Naubereit2018,Lawniczak2018,Bialous2016,Georgopulos1981,Biswas1992}, and thus our choice for this paper. For a recent discussion of this issue in a different context see Ref.~\cite{Naubereit2018}.

Due to its nature $P_k^\delta$ is a highly fluctuating quantity from a value of $k$ to the next for individual spectra. Moreover, in its natural logarithmic representation, the number of points is much larger for large values of $\log k$. This is not an important caveat when comparing with the GOE results or the Poisson results. However, when fitting for estimating $\varphi$ a direct least squares fit of the results is usually biased to give much more importance to large $k$ values as we have many more points in that region. This is partially compensated by the fact that deviations from GOE results tend to be larger in the low $k$ region but normally the bias is still there. So, when we cannot perform an ensemble average and to avoid this problem (and also for representation purposes) an average of $P_k^\delta$ over different values of $k$ is performed. Of course, there are different ways to perform this average in the literature which, up to now, have not been given a particular importance \cite{Molina2007,MurPetit2015}. We have explored this question in more detail and found that this process of averaging may induce systematic errors in the estimation of $\varphi$. Our findings are summarized in the Appendix \ref{Ap:methods}. Our final conclusion is that the optimal way of averaging is done in intervals of $k$ (and not $\log k$ as was done in some previous works) with a number of intervals $n_{int}$ that depends on the dimension $d$ and the value of $\varphi$. However, this choice is not very critical and can be estimated with a simple formula $n_{int}=0.5*d^{0.65}$ that gives a number that can be safely used for all values of $\varphi$ which is important as in practical cases $\varphi$ is, in principle, unknown. The results using $\iPk$ that we present in the following section are performed with this procedure.

\section{Results}
\label{sec:results}

In this section we present the results of the estimation of the fraction $\varphi$ of observed levels in individual spectra from fits to the equations \eqref{ec:D3phi} and \eqref{ec:Pkphi} for $\Delta_3$ and $P_k^{\delta}$, and the comparison of their efficiency in terms of accuracy and precision. We remind the reader that by accuracy we mean the closeness of the estimated value to the real value and by precision we refer to the statistical variability of the estimation. From now on we will call $\varphi^{\Delta}$ the estimation of $\varphi$ obtained from the fit to the expression \eqref{ec:D3phi} for $\Delta_3$, $\varphi^P$ the estimation of $\varphi$ obtained from the fit to the expression \eqref{ec:Pkphi} for $P_k^{\delta}$, and $\varphi^e$ the value of $\varphi$ initially set for the ensemble, to distinguish between them when necessary.

We have generated ensembles of 1500 GOE spectra for dimensions $d = 100$ to $2500$ 
and values of $\varphi^e$ from 0.3 to 1. Then for each ensemble we have proceeded as follows: i) calculate $\Delta_3$ and $P_k^{\delta}$ for each individual spectrum ii) perform fits to Eqs. \eqref{ec:D3phi} and \eqref{ec:Pkphi} to obtain the estimations of $\varphi$, and iii) calculate the mean and standard deviation of the 1500 estimations of $\varphi$ from the ensemble for each statistic. Finally we analyze the precision in terms of the standard deviations and the accuracy by comparing with the exact value of $\varphi^e$ which has been set for each ensemble. The actual distribution of the results of the estimation is different from a Gaussian, in part due to the fact that the value of $\varphi$ is bounded between $0$ and $1$. Nevertheless, the value of the standard deviation still is a good measure of the precision of the method used. However, for a full understanding of the meaning of the result 
an analysis of the full distribution of the estimated values of $\varphi$ is needed. Some examples will be provided after discussing the results for the mean and standard deviation.

\subsection{Results for spectra of dimension $d=2500$}

In Table \ref{tab:gaussN2500} we show the results of the mean $\mu$ and standard deviation $\sigma$ of the estimations of $\varphi$ for each statistic in this ensemble. For the fits with the $\Delta_3(L)$ we use values of $L$ up to $L=d/3.5$. We have checked that the results do not depend very much on the chosen limit as long as there are enough points to make a proper fit. We choose a number of intervals in the $k$-axis of $80$ for the fits with the $P_k^\delta$. It can be seen that the accuracy is quite good for the ensemble of estimations and for both statistics. The mean values are very close to the values of $\varphi^e$ for the corresponding ensemble, always within one sigma. The precision, as measured by $\sigma/\mu$, lowers as the fraction of observed levels decrease, as one would expect. However, in this quantity there are significant differences between $\Delta_3$ and $P_k^\delta$. While in the former case the precision goes from $1.4\%$ for $\varphi^e=0.95$ to about $52\%$ for $\varphi^e=0.3$, in the later case the precision goes from $1\%$ to $12\%$. We can then conclude that for this large dimensions a fit to the formula for the $P_k^\delta$ gives better results for the estimation of experimental spectra. 

\begin{table}
\caption{Results of the average and standard deviation of the fits for $\varphi$ ($\mu$ and $\sigma$) for a GOE ensemble of $1500$ matrices of dimension $d=2500$. The fits using $P_k^\delta$ were obtained with $80$ intervals in the $k$ axis while the fits using $\Delta_3$ were obtained with the results up to a maximum value of $L$ of $d/3.5$.}
\begin{tabular}{|c|c|c|c|c|}
\cline{2-5}
\multicolumn{1}{c|}{{}} & \multicolumn{2}{c|}{{$\overline{P_k^\delta}$}} & \multicolumn{2}{c|}{{$\overline{\Delta_3}$}} \\
\hline
{$\varphi$} & {$\mu$} & {$\sigma$} & {$\mu$} & {$\sigma$}  \\
\hline
\hline
0.3  & 0.311   & 0.037    & 0.33   & 0.17   \\
0.5  & 0.507   & 0.031     & 0.52   & 0.12      \\
0.6  & 0.603   & 0.027     & 0.618  & 0.094     \\
0.7  & 0.698   & 0.022     & 0.719  & 0.070     \\
0.75 & 0.746   & 0.019     & 0.764  & 0.059     \\
0.8  & 0.792   & 0.018     & 0.809  & 0.046     \\
0.85 & 0.839   & 0.016     & 0.856  & 0.035    \\
0.9  & 0.887  & 0.013     & 0.905  & 0.025    \\
0.95 & 0.935  & 0.010 &  0.952  & 0.013   \\
\hline
\end{tabular}
\label{tab:gaussN2500}
\end{table}

\begin{figure}[h]
\includegraphics[width=0.45\columnwidth]{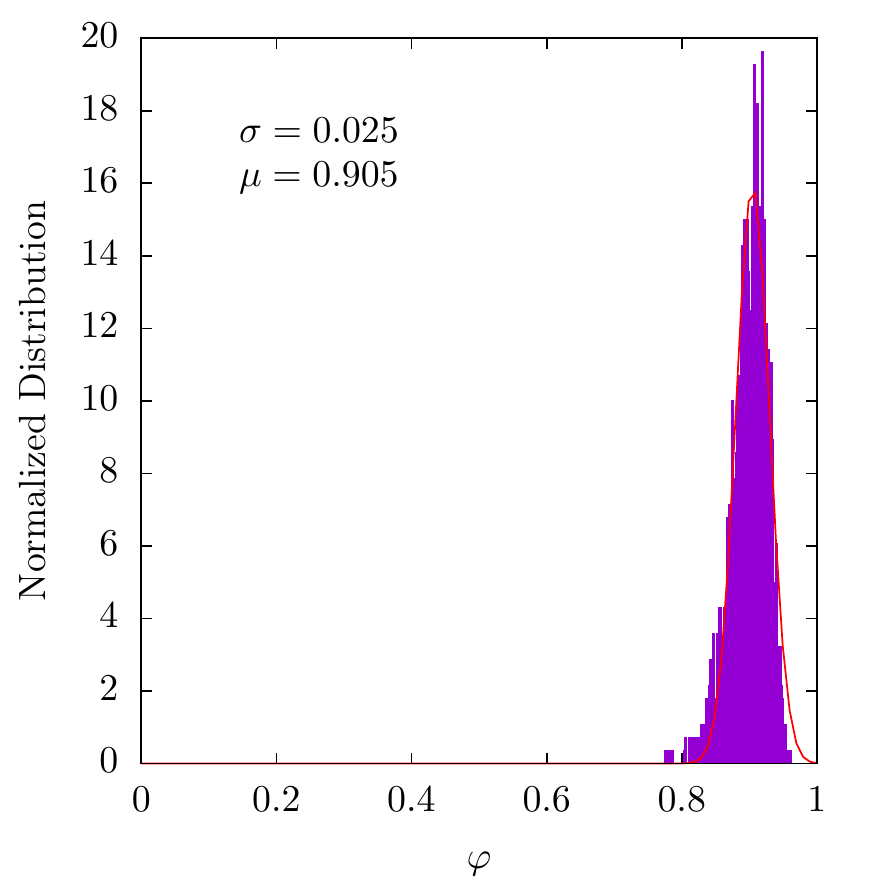}
\includegraphics[width=0.45\columnwidth]{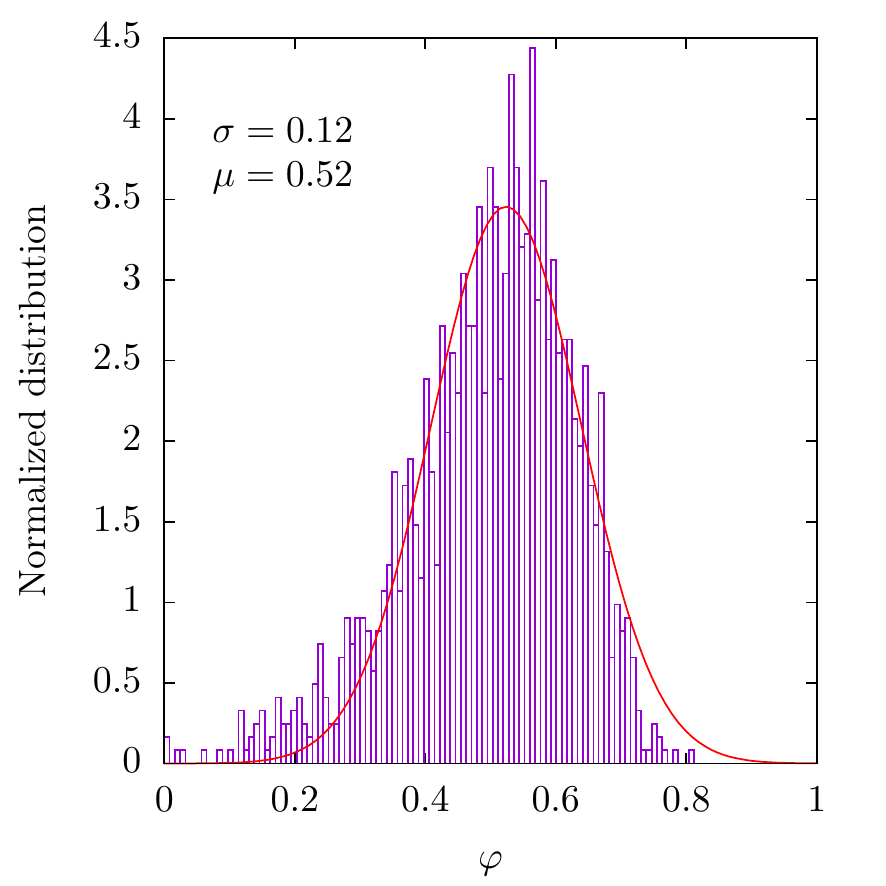}
\caption{Distribution of the estimations $\varphi^{\Delta}$ for an ensemble of 1500 GOE matrices of dimension $d=2500$ with $\varphi^e=0.9$ (left) and $\varphi^e=0.5$ (right). Notice that the axis are different between the figures. The solid red line is the best fit to a Gaussian distribution with the parameters shown in the figure.}
\label{fig:D3distN2500_90}
\end{figure}

\begin{figure}[h]
\includegraphics[width=0.45\columnwidth]{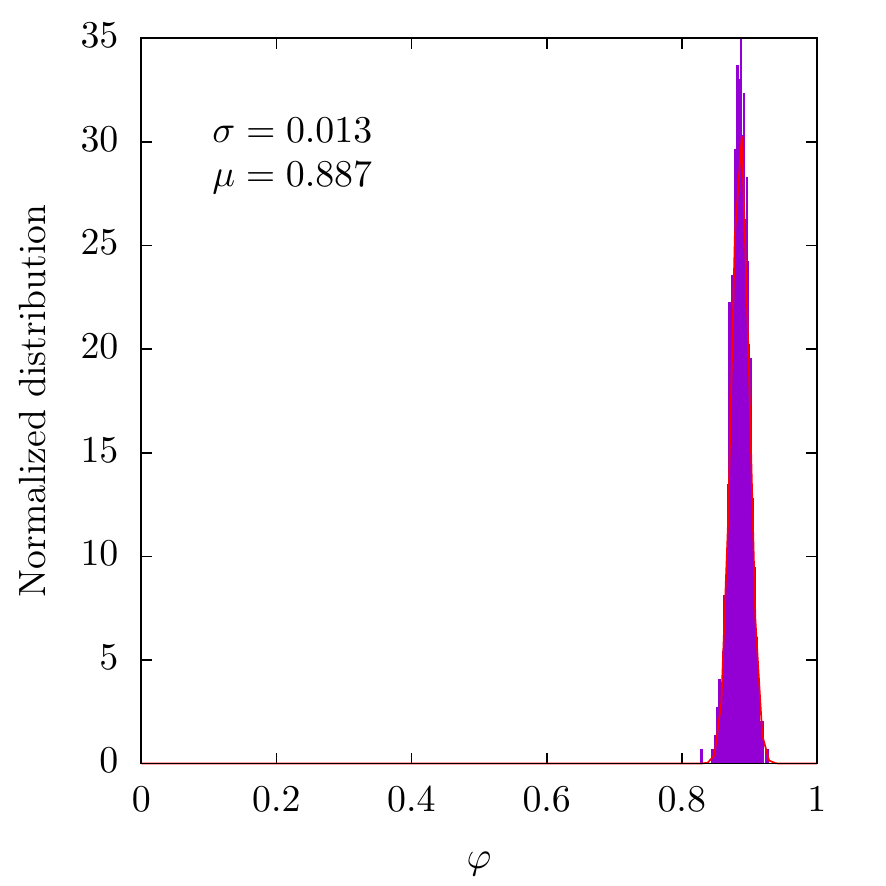}
\includegraphics[width=0.45\columnwidth]{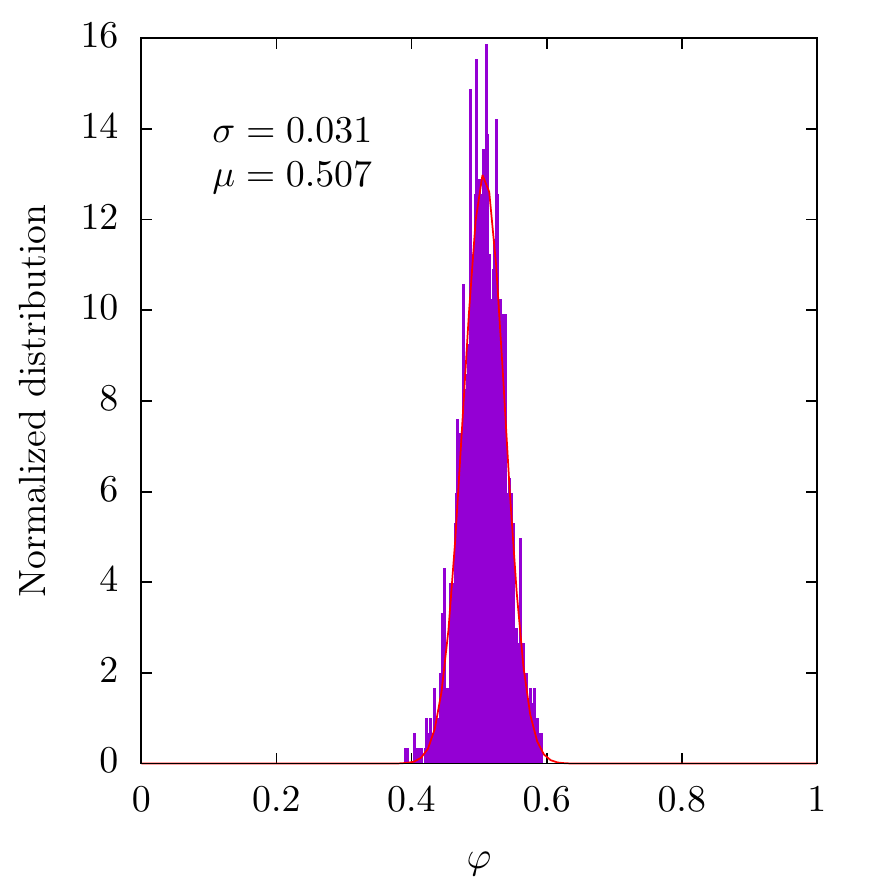}
\caption{Distribution of the estimation of $\varphi^P$ for the same ensembles as the previous figure. The solid red line is the best fit to a Gaussian distribution with the parameters shown in the figure. The x-axis are the same as in Fig. \ref{fig:D3distN2500_90} but not the y-axis.}
\label{fig:PkdistN2500_90}
\end{figure}

In Figs. \ref{fig:D3distN2500_90} and \ref{fig:PkdistN2500_90} we show the distribution of estimations $\varphi^P$ and $\varphi^{\Delta}$ for the ensemble with $\varphi^e=0.9$ and $\varphi^e=0.5$. Together with the histograms we have represented Gaussian curves with the corresponding parameters $\mu$ and $\sigma$ obtained for the ensemble from Table \ref{tab:gaussN2500}. The distributions have a reasonably Gaussian shape and the standard deviations can be used for estimating the dispersion of values of $\varphi$, although the full distribution is needed
for a correct interpretation of the fitting results. This is particularly evident in the case of the distribution of the results for the fitting with $\Delta_3$. The comparison of the normalized distribution of the estimation for the two different statistics clearly supports the choice of the fits of $P^{\delta}_k$ for the estimation of $\varphi$ for this dimension. This difference diminishes for lower dimensions, as we will see in the next section with $d=200$. 

\subsection{Ensembles of GOE spectra of dimension $d=200$}

\begin{figure}[h]
\includegraphics[width=0.45\columnwidth]{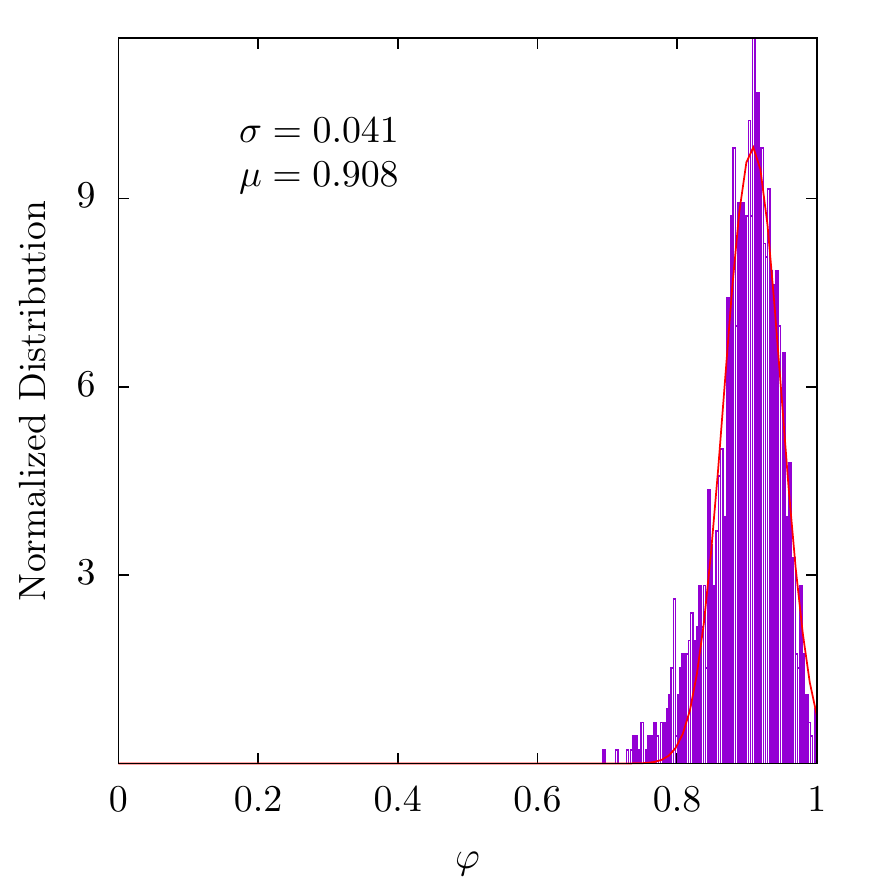}
\includegraphics[width=0.45\columnwidth]{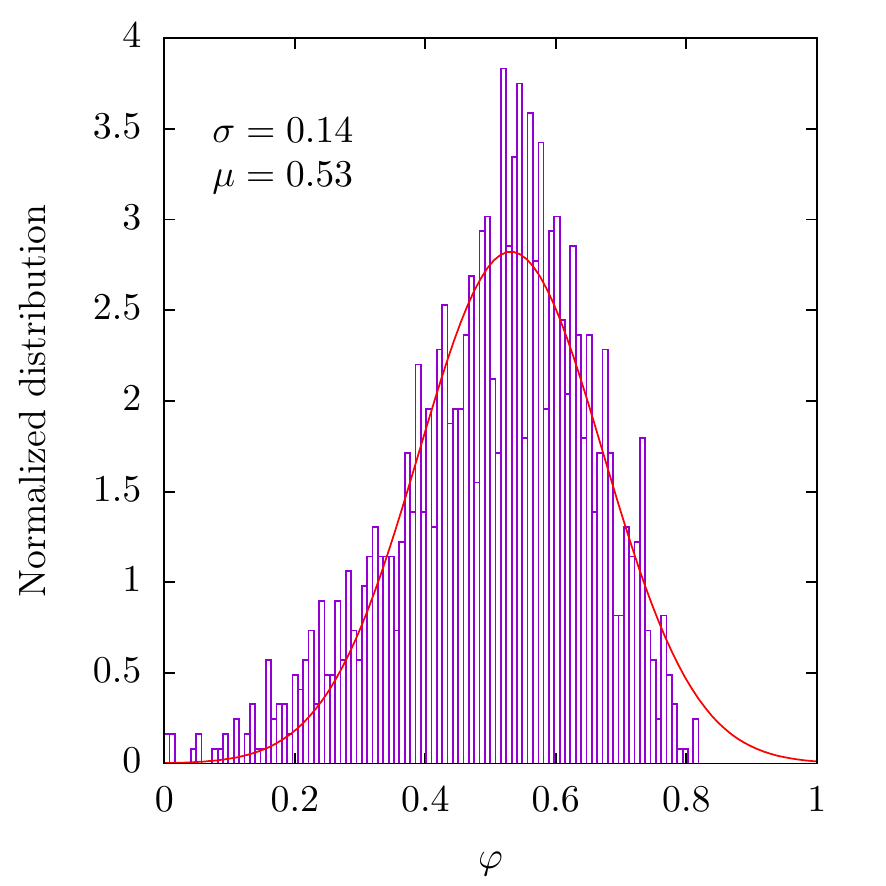}
\caption{Distribution of the estimations $\varphi^{\Delta}$ for an ensemble of 1500 GOE matrices of dimension $d=200$ with $\varphi^e=0.9$ (left) and $\varphi^e=0.5$ (right). Notice that the axis are different between the figures. The solid red line is the best fit to a Gaussian distribution with the parameters shown in the figure.}
\label{fig:D3distN200_90}
\end{figure}

\begin{figure}[h]
\includegraphics[width=0.45\columnwidth]{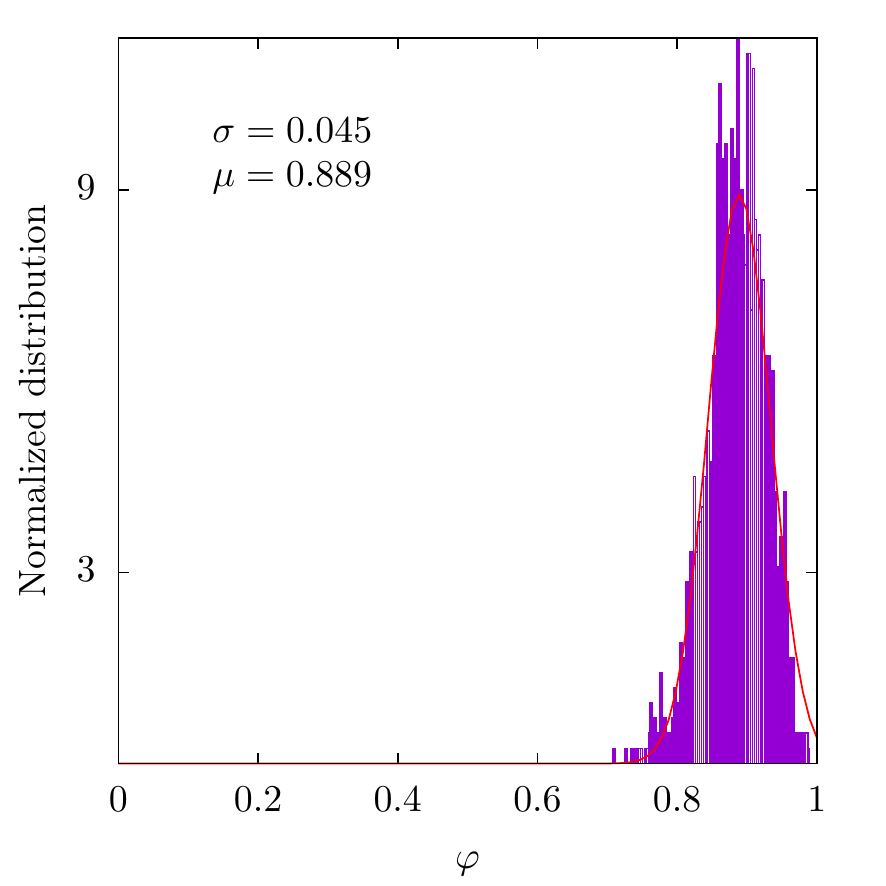}
\includegraphics[width=0.45\columnwidth]{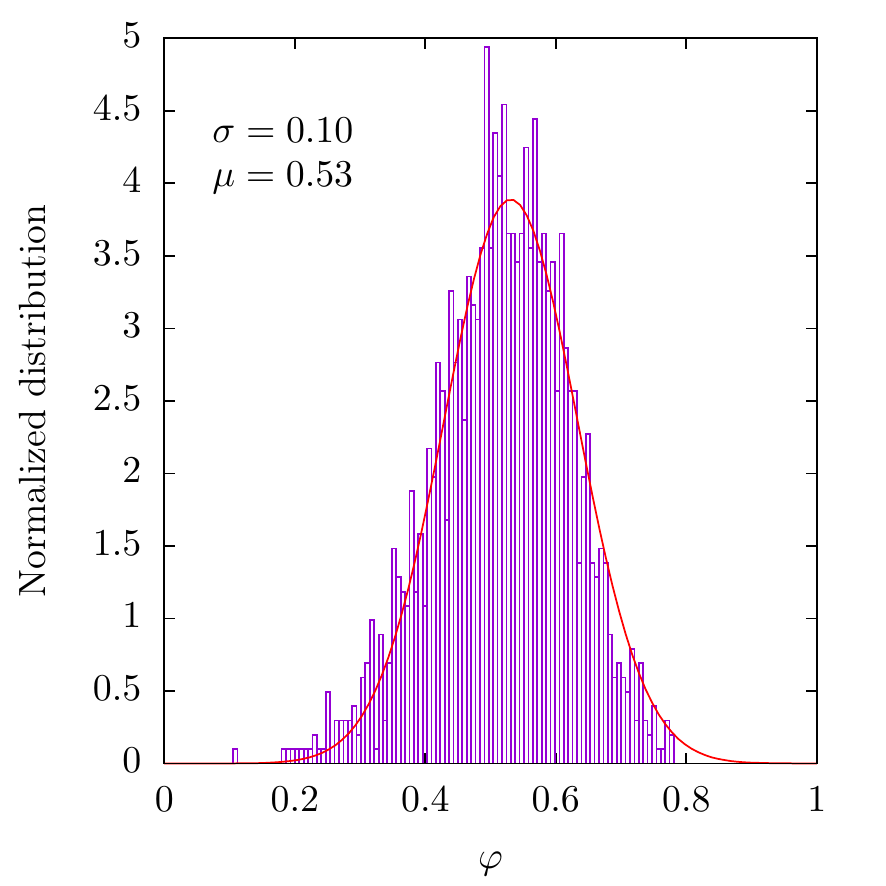}
\caption{Distribution of the estimation of $\varphi^P$ for the same ensembles as the previous figure. The solid red line is the best fit to a Gaussian distribution with the parameters shown in the figure. The x-axis are the same as in Fig. \ref{fig:D3distN200_90} but not the y-axis.}
\label{fig:PkdistN200_90}
\end{figure}

In Table \ref{tab:gaussN200} we show the results of the mean $\mu$ and standard deviation $\sigma$ of the estimations of $\varphi$ for each statistic in this ensemble, that is, the same as in Table \ref{tab:gaussN2500} for $d=2500$. By comparing both we can observe the same general trends: the mean values $\mu \pm \sigma$ are compatible with the values of $\varphi^e$ for the corresponding ensemble, though the precision ($\sigma/\mu$) is lower as the fraction of observed levels decrease. Here as well there are no significant differences between the estimations with $\Delta_3$ and with $P_k^\delta$ dealing with accuracy. There are still differences dealing with precision and better results for the estimations with $P_k^\delta$ are observed in the case of smaller $\varphi$ although the differences are smaller than in the case of $d=2500$. Another difference of Table \ref{tab:gaussN200} with respect to Table \ref{tab:gaussN2500} is the general increase in the standard deviations, as expected when the dimension decreases.
\begin{table}
\caption{Results of the average and standard deviation of the fits for $\varphi$ ($\mu$ and $\sigma$) for a GOE ensemble of $1500$ matrices of dimension $d=200$. The fits using $P_k^\delta$ were obtained with $18$ intervals in the $k$ axis while the fits using $\Delta_3$ were obtained with the results up to a maximum value of $L$ of $d/3.5$.}
\begin{tabular}{|c|c|c|c|c|}
\cline{2-5}
\multicolumn{1}{c|}{{}} & \multicolumn{2}{c|}{{$\overline{P_k^\delta}$}} & \multicolumn{2}{c|}{{$\overline{\Delta_3}$}} \\
\hline
{$\varphi$} & {$\mu$} & {$\sigma$} & {$\mu$} & {$\sigma$} \\
\hline
\hline
0.3  & 0.34  & 0.13   & 0.32   & 0.20  \\
0.5  & 0.53  & 0.10   & 0.53   & 0.14  \\
0.6  & 0.623 & 0.089  & 0.62   & 0.12  \\
0.7  & 0.700 & 0.081  & 0.714  & 0.086 \\
0.75 & 0.753 & 0.068  & 0.764  & 0.082 \\
0.8  & 0.804 & 0.060   & 0.808  & 0.068 \\
0.85 & 0.844 & 0.054  & 0.858  & 0.056 \\
0.9  & 0.889 & 0.045  & 0.908  & 0.041  \\
0.95 & 0.931 & 0.035  & 0.951  & 0.039  \\
\hline
\end{tabular}
\label{tab:gaussN200}
\end{table}
For a graphical comparison we show in Figs. \ref{fig:D3distN200_90} and \ref{fig:PkdistN200_90} the distribution of estimations $\varphi^P$ and $\varphi^{\Delta}$ for the ensemble with $\varphi^e=0.9$ and $\varphi^e=0.5$, that is, the same as in Figs. \ref{fig:D3distN2500_90} and \ref{fig:PkdistN2500_90} for $d=2500$. It can be seen as the distributions are more extended in this case and the shape differs more from Gaussian, specially in the case of the fitting with~$\Delta_3$.

Thus, in view of Table \ref{tab:gaussN200}, the main conclusion of the comparison between the estimations $\varphi^P$ and $\varphi^{\Delta}$ changes for this lower dimension: accuracy and precision are similar when using $\Delta_3$ or $P_k^{\delta}$ for larger values of $\varphi$ although $P_k^{\delta}$ is still much better for smaller values. In Appendix \ref{Ap:table1000} we also show results for an intermediate value $d=1000$ confirming this trend.

Now, in the next section we would like to give some guidance on how to proceed to obtain the best range estimation of $\varphi$ for a given single spectrum.

\subsection{A single spectrum: How to proceed}

\begin{figure}[h]
\includegraphics[width=0.4\textwidth]{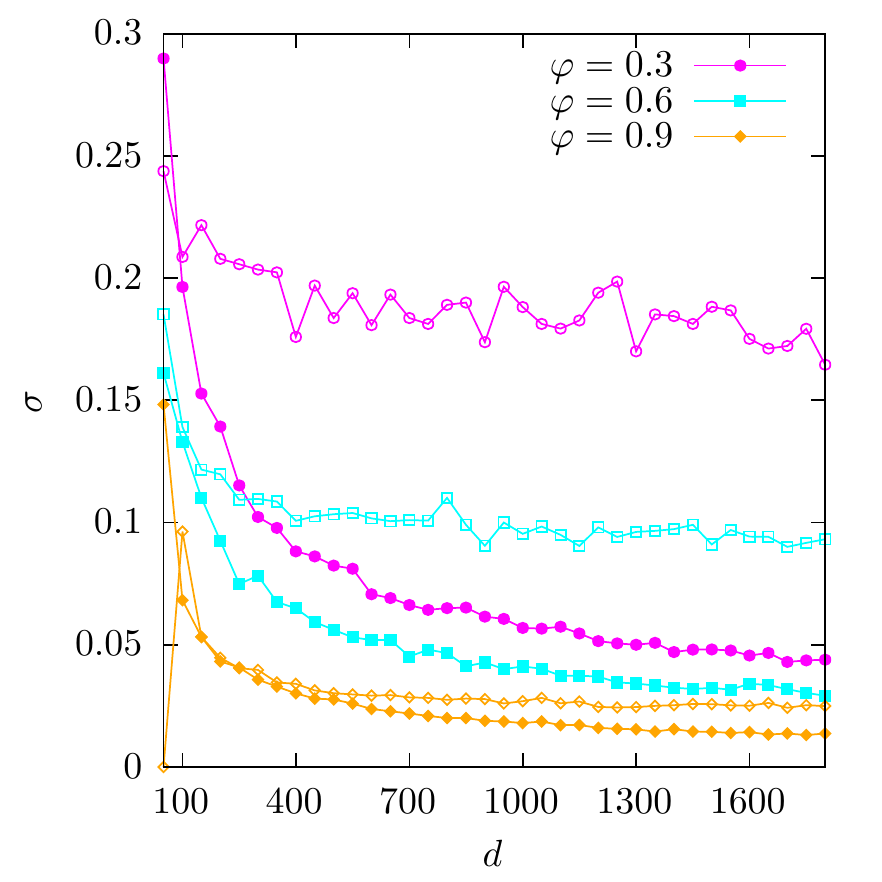}
\caption{Value of the standard deviation of the ensemble distributions of the estimations of $\varphi$ versus the dimension of the spectrum. Circles correspond to $\varphi^e=0.3$, squares to $\varphi^e=0.6$ and diamonds to $\varphi^e=0.9$. Empty symbols correspond to the estimations with $\Delta_3$ and filled symbols to the estimations with $P_k^\delta$}
\label{fig:sigma_vs_d}
\end{figure}

First, a completion of the results presented in the previous sections for a high and a low dimension is in order. We have perform analyses of ensembles of several dimensions and we found no significant differences in accuracy between $\iPk$ and $\iD$ but we found differences in precision. In Fig. \ref{fig:sigma_vs_d} we show the evolution of the standard deviation of the ensemble distributions of $\varphi$ values for $\iPk$ and $\iD$ for $\varphi^e=0.3, 0.6$ and 0.9. It can be seen that, except for the lowest dimensions, the precision of the fit with $\iPk$ is better and the difference is bigger as the fraction of observed levels decrease. We remind that good precision means that the probability to obtain a value of $\varphi$ for the single spectrum of study (a single member of the ensemble) belongs to a narrow interval around the average value. 

Now, when performing a fit with any of the two statistics, we can give the first conservative range estimation by checking to which of the ensembles our fitted value of $\varphi$ could belong in the tables \ref{tab:gaussN2500} ($d=2500$), \ref{tab:gaussN1000} ($d=1000$) and \ref{tab:gaussN200} ($d=200$) for a dimension similar to that of the spectrum of study (or making our own by a GOE Monte Carlo calculation like we described earlier). But the best range estimation of $\varphi$ should be obtained by calculating the ensembles of spectra of the exact dimension for numerically obtaining the full joint distribution of the fitted $\varphi$ and the actual $\varphi
^e$. From that numerical calculation we should obtain the conditional probability distribution given our result for the fitted $\varphi$, and the best range estimation from the standard deviation of this distribution.

So, although both statistics give meaningful results, we recommend to rely on $P_k^\delta$ specially for higher dimensions. Moreover, we also believe calculating $P_k^\delta$ is more ''friendly'' to the non-practitioner. Although, when we use $P_k^\delta$, we have to take into account that we have to choose the optimal number of intervals $n_{int}$ to perform the average in intervals of $k$, specially when ensemble averaging is not possible because of the small size of the spectrum. As explained in the Appendix \ref{Ap:methods}, we only have to use a simple rule to obtain safe results, that is, performing the average dividing in $n_{int}$ intervals the range $[k]$, with $n_{int}=0.5*d^{0.65}$. Though the optimal $n_{int}$ in principle depends both on the dimension and the value of $\varphi$, a further refinement of $n_{int}$ taking this into account would not produce such significant improvement to make worth the calculation of new ensembles, as the choice of $n_{int}$ is robust against ample variations, as explained in the Appendix. The formula for $n_{int}$ can be safely used when $\varphi$ is completely unknown, and when we have some hint about its value then the exponent can be chosen nearer to 0.6 for lower values of $\varphi$ and nearer to 0.7 for higher values of $\varphi$. 

\section{Conclusions}
\label{sec:conclusions}
In order to use RMT for estimating properties of chaotic experimental spectra it is imperative to understand the precision and accuracy of the statistics and the fitting procedure of choice as a function of the size of the spectra and other properties of the system. We have done a systematic analysis of the methods for estimating the fraction of observed levels $\varphi$ in chaotic spectra based on the analysis of long-range spectral correlations. In particular, we have studied the methods using the power spectrum of the $\delta_n$ statistic $P_k^\delta$ and the $\Delta_3$ of Dyson and Mehta. To analyze the methods we have prepared ensembles of GOE spectra taken out randomly a definite number of levels so $\varphi$ is known.

Both statistics give reasonable results, the accuracy of the estimations being similar and quite good even for relatively short sequences with many missing levels. Accuracy and precision are better, as expected, for larger sequences and larger values of $\varphi$. Our method of choice, however, would be to use $P_k^\delta$ as its precision is better, being the dispersion of the results smaller, specially for larger sequences. We remind that better precision for the result of the ensemble (lower standard deviations from the correct $\varphi$) implies less error in the 
estimation of the value of $\varphi$ from the fit
to a single spectrum of interest, as the value of $\varphi$ for any member of the ensemble would have more probability to lie in a narrow range of values around the correct one.

Moreover, in our opinion $P_k^\delta$ is more friendly and easier to interpret to scientists without experience in statistical analysis of spectra. It is based on a simple Fourier transform instead of being a particular definition in this field like the more complex $\Delta_3$. The caveat of using $P_k^\delta$ is that, when ensemble averages are not possible because of the small size of the spectrum, some extra averaging is needed for an unbiased estimation, but we have 
found a very simple rule which guarantees the safest results, that is, performing the average inside $n_{int}$ intervals of the range $[k]$, with $n_{int}$ given by formula \ref{eq:n_int}.

We have concentrated on the case where missing levels are missing randomly, for example, if the physical reason behind the missed observation of the levels is related to the nodal distribution of the wave functions. Similar Monte Carlo calculations can be used for making estimations of the number of missing levels when there are some correlations between the unobserved levels, for example, if the levels may be missed because near levels are not resolved experimentally as recent work have shown \cite{Lawniczak2018}.

We hope this work will contribute to the systematization of the use of long-range correlations for estimating the number of missing levels in experimental chaotic spectra, for quantum or wave systems. Without this type of analysis the statistical significance of any estimation based on the fit to a theoretical formula is unknown.

\appendix

\section{Method of calculation of $P_k^{\delta}$}
\label{Ap:methods}

The calculation of $P_k^{\delta}$ is straightforward, being a discrete Fourier transform of the function $\delta_n$ defined by Eq. \eqref{eq:deltan}. As seen in Fig. \ref{fig:form_ensembles_Pkphi} when calculating $P_k^{\delta}$ averaged over an ensemble one obtains sets of points which present a smooth behavior. 
When calculating $P_k^{\delta}$ for a single spectrum the result is not so smooth, so there is the need to smoothen it in some way to be able to see the trend more clearly, specially in cases like the one we approach in this work when we have to fit the formula to a set of points in order to estimate the fraction of observed levels $\varphi$.

When the spectrum of interest is reasonably large it is usual to divide it in several sequences and then perform averages exactly as for an ensemble of spectra.
When this is not possible there is still another usual method which consists in dividing the $x$ axis, that is, the range of frequencies $k$, into several intervals with a fixed length and represent only one point with the mean value for each interval. This can be done before or after taking the logarithm of the frequency, that is, we can divide the range $[k]$ or the range $[\log k]$. In this Appendix we present a comparison of both methods and, in view of the analysis and conclusions, we have selected the optimal method which we use for the results presented in the paper.

When possible, the optimal average to smooth the results is a two-fold one, that is, divide the spectrum in several sequences and perform a first average and then divide the result in several intervals and perform the second average as described. Here we analyze only the second method as the spectra of interest are typically small and we use spectra of dimension $d=200$. In Table \ref{tab:gaussN200AVGk} we show for comparison the fitting results for averaging in the range $[k]$ and in the range $[\log k]$, choosing the optimal value for the number of intervals in both cases. It can clearly be seen accuracy an precision are much better when averaging in the range $[k]$. We have also seen these trends in the rest of the cases of different dimensions analyzed, that is, an overestimation of the value of $\varphi$ and an increase in the values of $\sigma$ when averaging in the range $[\log{k}]$. Only for $\varphi
^e=0.95$ the results are better for averaging in the range of $[\log k]$, although in this case the results are actually very far from Gaussians and the fit in many cases results in $\varphi=1$.

\begin{figure}[h]
\includegraphics[width=\columnwidth]{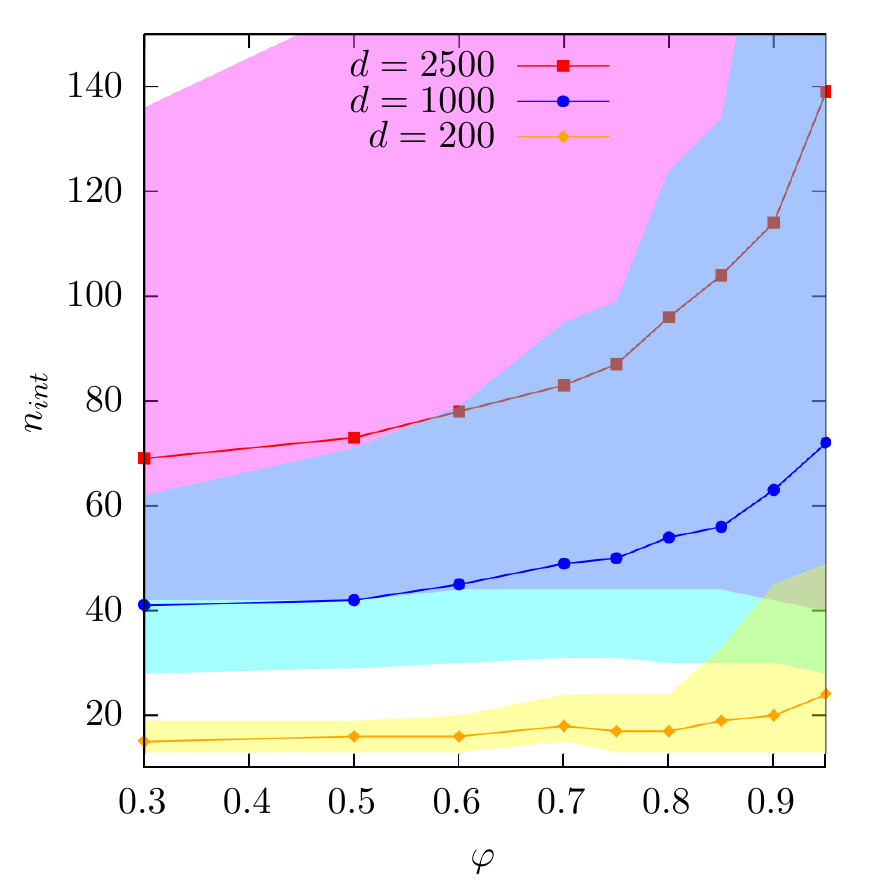}
\caption{Optimal number of intervals for fitting $\varphi$ using $\iPk$ versus $\varphi$ for different values of the dimension. The transparent shaded areas encompass all the values of $n_{int}$ for which the systematic error in the average estimation of $\varphi$ is less than 0.05. That of $d=2500$ reaches values above 400 for the highest values of $\varphi$, but the $y$ axis has been set up to 150 to see the three cases more clearly.}
\label{fig:optimal-vs-phi}
\end{figure}

\begin{figure}[h]
\includegraphics[width=\columnwidth]{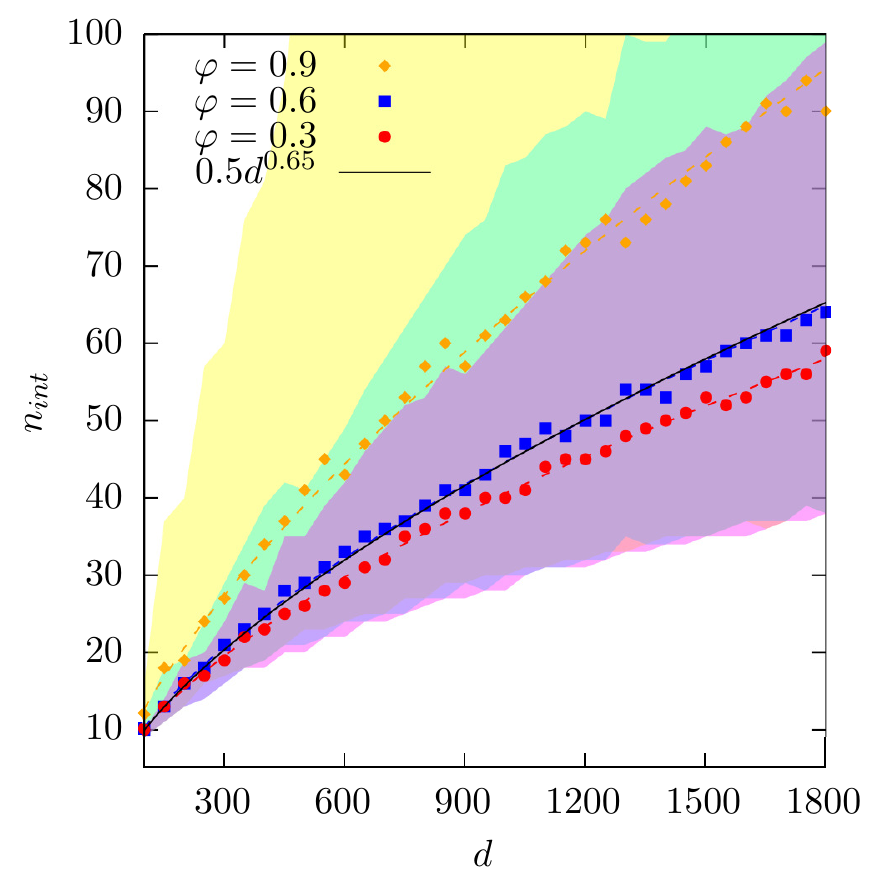}
\caption{Optimal number of intervals for fitting $\varphi$ using $\iPk$ versus the dimension for $\varphi=0.3, 0.6$ and 0.9. The black solid line corresponds to the expression \ref{eq:n_int} and the dashed curves correspond to the fit to equation \ref{eq:n_int_fit}. The transparent shaded areas encompass all the values of $n_{int}$ for which the systematic error in the average estimation of $\varphi$ is less than 0.05. For $\varphi=0.9$ it reaches values above 300 for the highest values of $d$, but the $y$ axis limit has been set up to 100 to see the three cases more clearly.}
\label{fig:optimal-vs-dim}
\end{figure}

\begin{table}
\caption{Ensemble of 1500 GOE matrices of dimension $d=200$. Comparison of results for $\mu$ and $\sigma$ with $10$ intervals averaged on $\log k$ (the best result in this case) and $18$ intervals averaged on $k$.}
\begin{tabular}{|c|c|c|c|c|}
\cline{2-5}
\multicolumn{1}{c|}{{}} & \multicolumn{2}{c|}{{$\overline{P_k^\delta}$} average on $\log{k}$} & \multicolumn{2}{c|}{{$\overline{P_k^\delta}$} average on $k$ } \\
\hline
{$\varphi$} & {$\mu$} & {$\sigma$} & {$\mu$} & {$\sigma$} \\
\hline
0.3  & 0.45  & 0.20  & 0.34   & 0.13  \\
0.5  & 0.63  & 0.15  & 0.53    & 0.10   \\
0.6  & 0.71  & 0.13  & 0.623    & 0.089   \\
0.7  & 0.79 & 0.10 & 0.700    & 0.081  \\
0.75 & 0.832 & 0.090 & 0.753    & 0.068  \\
0.8  & 0.872 & 0.074 & 0.804    & 0.060  \\
0.85 & 0.910 & 0.067 & 0.844    & 0.054  \\
0.9  & 0.953 & 0.051 & 0.889    & 0.045  \\
0.95 & 0.964 & 0.024 & 0.931 & 0.035  \\
\hline
\end{tabular}
\label{tab:gaussN200AVGk}
\end{table}
We have also analyzed the fitting results depending on the number of intervals $n_{int}$ and the dimension. For each dimension and each value of $\varphi^e$ of the ensembles generated we have perform the calculations to obtain the distributions of values of $\varphi$ using different numbers of intervals and calculating the systematic error $|\mu-\varphi^e|$ in each case. We call the {\it optimal} $n_{int}$ the one which minimizes this error. We have seen that using a large number of intervals or a small number of intervals produces systematic deviations in the fitting of $\varphi$ as the weight of the large and low $k$ sections in the least squares procedures varies. Except for $\varphi<0.5$ there is a wide optimal number of intervals which is plotted in Fig. \ref{fig:optimal-vs-phi} for different dimensions. The shaded area represents the number of intervals where there is less than 0.05 of systematic error in the average estimation of $\varphi$. 
In Fig. \ref{fig:optimal-vs-phi} it can also be seen that the dependence of $n_{int}$ on the value of $\varphi$ is neither very critical, being also a wide optimal range of $n_{int}$ common to all the $\varphi$ values. For example, for $d=200$ we have chosen $n_{int}=18$ as the optimal number of intervals common to all the values of $\varphi$. But in order to better see this dependence we have also represented $n_{int}$ versus the dimension for three different values of $\varphi$ in Fig. \ref{fig:optimal-vs-dim}. We have found a simple functional form which describes the dependence very well: 
\begin{equation}
    n_{int} = a(\varphi)*d^{b(\varphi)}
    \label{eq:n_int_fit}
\end{equation}
where the parameters $a$ and $b$ in principle depend on the value of $\varphi$, but again in this figure we can clearly see that this dependence is not very critical as the choice is robust against ample variations: inside the shaded area the error in the average estimation of $\varphi$ is less than 0.05. Thus, if $\varphi$ is completely unknown one could safely choose the expression:
\begin{equation}
    n_{int} = 0.5*d \; ^{0.65}
    \label{eq:n_int}
\end{equation}
represented by the black solid line in the figure.
It can be seen that we would have obtained good results with this choice if $\varphi$ would have been completely unknown as the black solid line is inside all the shaded areas. The fitting parameters for the three values of $\varphi$ are shown in Table \ref{tab:fits-n_int}.
When we have some hint about the estimation of $\varphi$ then we would choose a value for the parameter $b$ nearer to 0.6 for lower $\varphi$ or nearer to 0.7 for higher $\varphi$.

\begin{table}
  \begin{tabular}{|c||c|c|}
  \hline
  {$\varphi$} & $a$ & $b$ \\
  \hline
  0.3  & 0.61 & 0.61 \\
  0.6  & 0.55 & 0.64 \\
  0.9  & 0.51 & 0.70 \\
  \hline
\end{tabular}
\caption{Fitting parameters $a$ and $b$ in Eq. \ref{eq:n_int_fit} for several values of $\varphi$}
\label{tab:fits-n_int}
\end{table}

To summarize this systematic study on the average to be performed to obtain the best estimation of $\varphi$ from the fit to $P_k^{\delta}$, we found essentially two rules: 1) Divide in a number of intervals the range $[k]$ (and not the range $[\log{k}]$) 2) Choose the number of intervals $n_{int}$ according to formula \ref{eq:n_int}. 

\section{Spectra of dimension $d=1000$}
\label{Ap:table1000}
We show in this section a table with the results for intermediate dimension $d=1000$, Table \ref{tab:gaussN1000}, in agreement with the trend explained in the main text.
\begin{table}[h]
\caption{Results of the average and standard deviation of the fits for $\varphi$ ($\mu$ and $\sigma$) for a GOE ensemble of $1500$ matrices of dimension $d=1000$. The fits using $P_k^\delta$ were obtained with $50$ intervals in the $k$ axis while the fits using $\Delta_3$ were obtained with the results up to a maximum value of $L$ of $d/3.5$.}
\begin{tabular}{|c|c|c|c|c|}
\cline{2-5}
\multicolumn{1}{c|}{{}} & \multicolumn{2}{c|}{{$\overline{P_k^\delta}$}} & \multicolumn{2}{c|}{{$\overline{\Delta_3}$}} \\
\hline
{$\varphi$} & {$\mu$} & {$\sigma$} & {$\mu$} & {$\sigma$} \\
\hline
\hline
0.3  & 0.329 & 0.058  &  0.33  &  0.18 \\
0.5  & 0.521 & 0.048  &  0.52  &  0.12 \\
0.6  & 0.613 & 0.043  & 0.620  & 0.097 \\
0.7  & 0.705 & 0.034  & 0.717  & 0.075 \\
0.75 & 0.751 & 0.031  & 0.761  & 0.065 \\
0.8  & 0.798 & 0.028  & 0.809  & 0.052 \\
0.85 & 0.845 & 0.023  & 0.857  & 0.040 \\
0.9  & 0.890 & 0.019  & 0.904  & 0.028 \\
0.95 & 0.937 & 0.014  & 0.952  & 0.016 \\
\hline
\end{tabular}
\label{tab:gaussN1000}
\end{table}

\newpage

\acknowledgements
We acknowledge financial support from Projects No. PGC2018-094180-B-I00 (MCIU/AEI/FEDER, EU) and No. RTI2018-098868-B-I00 (MCIU/AEI/FEDER, EU)
and CAM/FEDER Project No.\ S2018/TCS-4342 (QUITEMAD-CM). This research has been also supported by CSIC Research Platform on Quantum Technologies PTI-001.

\end{document}